\documentclass{llncs}

\usepackage{setspace}
\usepackage{textcomp}
\usepackage[english]{babel}
\usepackage[latin1]{inputenc}
\usepackage{verbatim}
\usepackage{fancyvrb}
\usepackage{graphics}
\usepackage{graphicx}
\usepackage{algorithm}
\usepackage{algpseudocode}
\usepackage{url}
\usepackage{cite}
\usepackage{stfloats}
\usepackage{longtable}
\usepackage{multirow}
\usepackage{hhline}
\usepackage{makecell}
\usepackage[table,xcdraw]{xcolor}
\usepackage{color}
\usepackage[caption=false]{subfig}

\def\ONT{\textsc{OASIS}}

\title{Blockchains through ontologies: the case study of the Ethereum ERC721 standard in \ONT{} (Extended Version)}

\author{Giampaolo Bella,\inst{1}   Domenico Cantone,\inst{1}  Cristiano Longo,\inst{2} \\ Marianna Nicolosi Asmundo,\inst{1}  Daniele Francesco Santamaria\inst{1}}

\institute{University of Catania\\
Department of Mathematics and Computer Science\\
Viale Andrea Doria, 6 - 95125 - Catania, ITALY\\ \email{\{giampaolo.bella, domenico.cantone,\\ marianna.nicolosiasmundo, daniele.santamaria\}@unict.it }\and The Sicilian Wheat Bank S.p.A, \email{longo@dmi.unict.it}}

\begin{document}

%\doublespacing

\maketitle

\begin{abstract}
Blockchains are gaining momentum due to the interest of industries and people in \emph{decentralized applications} (Dapps), particularly in those for trading assets through digital certificates secured on blockchain, called tokens. As a consequence, providing a clear unambiguous description of any activities carried out on blockchains has become crucial, and we feel the urgency to achieve that description at least for trading. This paper reports on how to leverage the \emph{Ontology for Agents, Systems, and Integration of Services} (``\ONT{}'') as a general means for the semantic representation of  smart contracts stored on blockchain as software agents. Special attention is paid to non-fungible tokens (NFTs), whose management through the ERC721 standard is presented as a case study.
\end{abstract}

\section{Introduction}
\label{sect:intro}

The last decade reports vast interest on blockchain technology and related applications from various realms, including economic, social, business, and academic ones. Beyond the financial speculation concerning  cryptocurrencies, the interest in blockchain technologies is mainly motivated by the fact that they realize decentralized and publicly shared ledgers, where third-party intermediaries demanding the client's total and unquestioned trust are no longer required. Blockchain technologies~\cite{christidis16} were precisely introduced to allow users to interact and run programs in a distributed way without the requirement of trusted entities, yet guaranteeing ownership, transparency, traceability, availability, continuity, and immutability of  digital shared assets. %This is  particularly interesting when the ledger is applied for managing legal issues in digital environments. 
Applications of  blockchain technologies range from the \emph{Internet of Things} (IoT) and robotics~\cite{xu2017taxonomy}, to commerce, healthcare, insurance, energy, laws, and communications.

One of the most popular applications of Turing-complete blockchains such as Ethereum~\cite{Szabo97} is the \emph{smart contract}. Smart contracts are self-executing and immutable   programs, autonomously running and verified on a distributed and decentralized public network, which implement decentralized applications on blockchain systems called \emph{Dapps}. In 2020, Dapps have particularly grown  as an exchange tool for non-fungible tokens (NFTs), namely digital certificates stored on the blockchain representing predetermined rights on certain unique assets. NFTs are mainly used as a proof of ownership of physical or digital products. Such tokens are routinely exchanged by users to witness that assets whose uniqueness is hard to demonstrate (for example, digital images) are owned in an exclusive way.  At the end of 2020, the market capitalization of NFTs reached the amount of 338 millions of U.S. dollars.\footnote{https://www.statista.com/statistics/1221742/nft-market-capitalization-worldwide/ (last access: 08/07/2021).} 
However, one of the main limitations of blockchains is the hard-coded nature of the transactions stored on them. As a consequence, it is hard to probe a blockchain, for example, to find smart contracts trading specific NFTs that satisfy certain requirements in terms of quality or quantity.

\begin{sloppypar}
Therefore, a formal semantic knowledge representation capturing the blockchain smart contracts as well as the activities carried out on it
facilitates the understanding of blockchain concepts, the interlinking with other out-of-chain information, and also formal reasoning. Moreover, a semantic conception of blockchains enables the automatic discovery of smart contracts, the interconnection of services running on different blockchains (i.e., cross-chain integration) and the integration between on-chain and off-chain services. These features turn out to be more interesting when smart contracts are implemented as mechanisms for generating and exchanging tokens. A desirable feature of token exchange systems is a precise and intelligent query mechanism capable
of determining what, when, and how certain assets have been generated, exchanged or destroyed. For example, intelligent systems may be aware of the activation of smart contracts for generating tokens with specific characteristics, e.g., of the type of exchanged asset, of the exchange of particular tokens at certain conditions, or of their destruction. More in general, intelligent systems may be aware of the activation of smart contracts and of all the related activities over the blockchain. %Because of blockchain transactions are hard-coded, deploying such a query engine turns out to be very difficult, at least on a general scale, even if the source code of the smart contracts that they induce is made publicly available.
\end{sloppypar}

Beyond a semantic representation of transactions and information stored in blocks, a real, semantically represented blockchain is effectively achievable if smart contracts are interpreted as reactive agents operating on a common environment, with a fully specified semantics of available operations,  committed actions, and stored data. 

The representation of blockchain actors requires ontological capabilities for fully representing agents and their interactions in a detailed way. This paper adopts the \emph{Ontology for Agents, Systems, and Integration of Services} (``\ONT{}'') towards the full, semantic representation of the Ethereum blockchain and the smart contracts running on it, with a special focus on the smart contracts compliant with the ERC721 standard\footnote{https://ethereum.org/it/developers/docs/standards/tokens/erc-721/} for NFTs management.

%and in particular, in the Ethereum blockchain, thus enabling an interconnected on-chain and out-of-chain public ledger.  

The paper is organized as follows. Section \ref{sec:related} presents related work. Section \ref{sec:prelim} outlines \ONT{}, whereas Sections \ref{sec:onto} and \ref{sec:erc} depict the ontology implementing the \ONT{} representation of, respectively,  the  Ethereum blockchain and the ERC721 standard. Finally, Section \ref{sec:conclusion} draws some conclusions and delineates future research directions.

\section{Related works} \label{sec:related}

Interest in symbiotically combining semantic web technologies and blockchains  is quite recent \cite{English2015, CanoCimmino19}. One of the areas of
investigation concentrates in developing a characterization of blockchain concepts and technologies through ontologies and of blocks and transactions meta-data. An ontological albeit theoretical approach at blockchain representation exists~\cite{Kruijff2017}.
Ruta \emph{et al.} propose   a blockchain framework for \emph{semantic web of things} (SWoT) contexts settled as a \emph{Service-Oriented Architecture} (SOA), where nodes can exploit smart contracts for registering, discovering, and selecting annotated services and resources~\cite{RSICPD18}.

Blockchain technologies are also exploited as a secure and public storage system for small data, including linked data, and to realize a more resilient architecture for the Semantic Web \cite{English2015}.

Other works aim at representing ontologies within a blockchain context. In \cite{KimL18}, ontologies are used as common data format for blockchain-based applications such as the proposed provenance traceability ontology, but are limited to implementation aspects of the blockchain. 

Fill discusses blockchains applied for tracking the provenance of knowledge, for establishing delegation schemes, and for ensuring the existence of patterns in formal conceptualizations using zero-knowledge proofs~\cite{Fill2019ApplyingTC}.  

A semantic interpretation of smart contracts as services bases on the \emph{Ethon} ontology \cite{ethon} exist~\cite{baqa2019}. The main limitation of that approach is the poor semantic description of smart contracts, thus hindering the discovery of unknown smart contracts and of the related operations fulfilled during their life-span.

\begin{sloppypar}
The \emph{Blockchain Ontology with Dynamic Extensibility} (BLONDiE) project \cite{Rojas2017AMP} provides a comprehensive vocabulary that covers the structure of different components (wallets, transactions blocks, accounts, etc.) of blockchain platforms (Bitcoin and Ethereum)  and that can be easily extended to other alternative systems. 
\end{sloppypar}

Finally, in  \cite{idc2021} the authors illustrated how the ontology \ONT{} is applied for ontologically describing digital contracts  (called \emph{ontological smart contracts}), intended as agreement among agents, and how they can be secured on Ethereum. 

In this contribution, the definition of digital contracts in  \cite{idc2021} is generalized to include blockchain smart contracts, intended as programs running on the blockchain and interpreted as digital agents in the \ONT{} fashion.
\section{Preliminaries} \label{sec:prelim}
The \emph{Ontology for Agents, Systems, and Integration of Services} (``\ONT{}'') \cite{woa2019} is an OWL 2 ontology  for representing agents and their activities. On one hand, the ontology  models (web) agents and, in particular, the way they interact and operate in a collaborative environment, regardless of the framework and language adopted for their implementation. Agents are mainly represented by means of the mentalistic notion of agent behavior inspired by \cite{tropos}, encompassing  goals and tasks that are achievable (either publicly available or exposed on request) by the agent, together with actions, sensors, and actuators used to perform operations.  On the other hand, \ONT{} is used to define actions that may be requested to other agents and their related  information such as operation inputs and outputs. Such requests are submitted by exchanging suitable fragments of \ONT{}, whereas agents whose capabilities are compatible with the requested actions are discovered by means of SPARQL queries performed over their behaviors. \ONT{} was applied to build a TRL3 prototype of a home assistant that
activates and manages applications, devices, and users interacting with each other within the environment~\cite{woa2019}. \ONT{} was also used to define agent agreements and store these in the IPFS file system \cite{ipfs} in order to reduce the transaction data stored directly on blockchain~\cite{idc2021}.

\begin{figure}[H]
    \centering
    \includegraphics[scale=0.6]{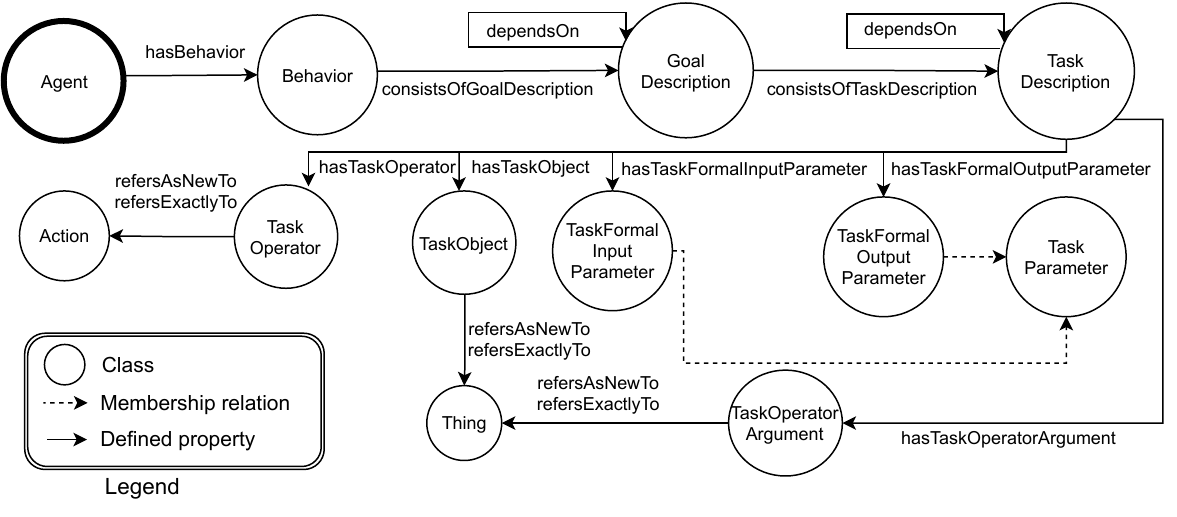}
       \caption{Agent representation in \ONT{}} \label{fig:oasis}
\end{figure}

OASIS models agents by publicly representing their behaviors. By exposing behaviors, agents report to the communication peers the set of operations that they are able to
perform and, eventually, the type of data required to execute these and the expected output. The representation of agents and their interactions in OASIS is carried out along three main steps. The first step consists in \textit{defining the template} for agent behaviour: templates are high-level descriptions of behaviors of abstract agents that can be implemented to define more specific and concrete behaviors of real agents. For example, a template may be designed for agents to  sell and ship products to buyers, and it may be implemented by any phone seller that ships its products using the Fedex courier: templates are useful to guide developers to define the most suitable representation of their agents. The second step consists in \textit{implementing a template},  %representing the agent behavior either by  or by defining it from scratch.
which requires a specification of the full operational details about the sought behavior. 

As depicted in Fig.~\ref{fig:oasis}, agent behaviors are represented by the goals to achieve, which in turn are related with their constitutional elements, namely tasks.
Tasks represent atomic operations that agents execute, and are described by the actions to be performed. Actions are drawn from a shared and common vocabulary, and can be simple or composed, eventually associated with requested input parameters and expected outputs. Finally, agent behaviors in \ONT{}  may be associated with conditionals \cite{idc2021}, adding constraints on the execution of actions and ensuring that certain conditions are verified before or after a task is executed.  \ONT{} conditionals are  OWL sentences that have the fashion of  \emph{Semantic Web Rule Language} (SWRL) rules \cite{swrl}, describing operations that must be triggered when certain conditions hold.

\section{Representing Ethereum through \ONT{} } \label{sec:onto}

\begin{figure}[H]
        \centering
        \includegraphics[scale=0.63]{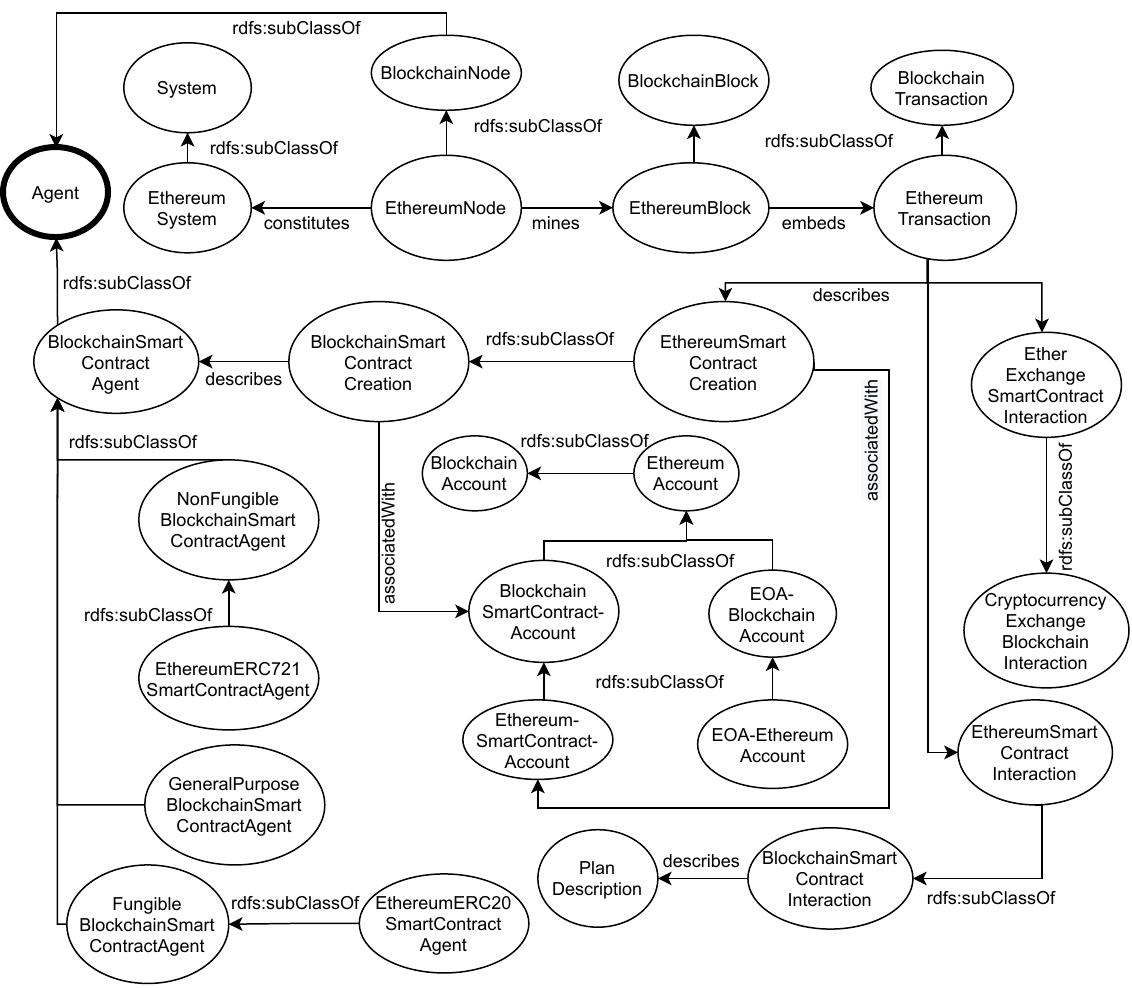}
        \caption{Representing the Ethereum blockchain  in \ONT{}}
        \label{fig:blockOnto}
\end{figure}

\begin{sloppypar}
In this section, we describe how the Ethereum blockchain is modelled in \ONT{}.\footnote{The ontology is reachable at \\ https://www.dmi.unict.it/santamaria/projects/oasis/sources/ether-oasis.owl} \ONT{} provides a different representation of blockchains with respect to Ethon \cite{ethon} and BLONDiE \cite{Rojas2017AMP}, since  the description of blockchain has to be aligned with the definitions of agent and agent behavior.  
\end{sloppypar}
\begin{sloppypar}
Ethereum is represented in \ONT{} by following the schema illustrated in Fig.~\ref{fig:blockOnto}. Ethereum blocks embedding transactions are represented by  instances of the class \textit{EthereumBlock} (subclass of \textit{BlockchainBlock}) and connected to the transactions contained in them by means of the object-property \textsf{embeds}. Each Ethereum transaction is identified by an instance of the class \textit{EthereumTransaction} (subclass of \textit{BlockchainTransaction}), encapsulating all the transaction information.

Block miners are identified by instances of the class \textit{EthereumNode} (subclass of \textit{BlockchainNode}) and linked to the mined block through the object-property \textsf{mines}: instances of \textit{BlockchainNode} are also instances of the class \textit{Agent}, representing  agents and provided with a behavior as in the \ONT{} fashion. Moreover, instances of \textit{BlockchainNode} (resp., \textit{EthereumNode}) are connected to instances of the class \textit{System} (resp., \textit{EthereumSystem}) by means of the object-property \textsf{constitutes}. Such a characterization of nodes, blocks, and transactions is the main difference with analogous approaches such as Ethon and BLONDiE, since it allows one to describe activities carried out by both in-chain and out-of-chain agents, thus providing a higher-level model of the two ecosystems and a means to unify them. Specifically, Ethereum activities are classified into two main categories, namely,  deployments of smart contracts, represented by instances of the class \textit{EthereumSmartContratCreation} (subclass of \textit{BlockchainSmartContractCreation}), and interactions with smart contracts, represented by instances of the class \textit{EthereumSmartContractInteraction} (subclass of the \textit{BlockchainSmartContractInteraction}). 
\end{sloppypar}

In \ONT{}, smart contracts deployed into the Ethereum blockchain  correspond to agents with well-defined behaviors: interactions with smart contracts are represented by \ONT{} \emph{plan executions} and linked to the behavior that induced the action. Specifically, a smart contract creation is represented by an instance of the class \textit{BlockchainSmartContractCreation}, which is related with the description of the agent describing its behavior by the object-property \textsf{describes}, the latter represented by an instance of the class \textit{BlockchainSmartContractAgent} (subclass of the class \textit{Agent}). Instances  of \textit{BlockchainSmartContractCreation} are also associated by means of the object-property \textsf{associatedWith} with the related Ethereum accounts, represented by instances of the class \textit{EthereumSmartContractAccounts} (sublcass of the class \textit{BlockchainSmartContractAccount}, which, in turn, is a subclass of \textit{BlockchainAccount}). Users are instead associated with  \emph{Ethereum externally owned accounts} (EOA) represented by instances of the class \textit{EOA-EthereumAccount} (sublcass of \textit{EOA-BlockchainAccount}, which is a subclass of \textit{BlockchainAccount}).

\begin{sloppypar}
\ONT{} identifies four main general categories of smart contract agents: a) smart contracts providing non-fungible token exchange mechanisms compatible with the Ethereum standard ERC721, which are represented by instances of the class \textit{EthereumERC721SmartContractAgent} (subclass of the class \textit{NonFungibleBlockchainSmartContractAgent}); b) smart contracts providing fungible token exchange mechanisms compatible with the Ethereum standard ERC20, which are represented by instances of the class \textit{EthereumERC20SmartContractAgent} (subclass of the class \textit{FungibleBlockchainSmartContractAgent}); c) agents responsible for exchanging Ether cryptocurrency, which are represented by instances of the class \textit{EtherExchangeSmartContractAgent} (subclass of the class \textit{CryptocurrencyExchangeBlockchainSmartContractAgent}); d) general purpose and user-defined smart contract agents that do not enjoy the characteristics of the aforementioned smart contracts, which are introduced by instances of the class \textit{GeneralPurposeBlockchainSmartContractAgent}.
\end{sloppypar}

\begin{sloppypar}
In \ONT{}, agents may perform actions autonomously or as response to requests of executing some operations submitted by a peer. Concerning the blockchain ecosystem, we mainly limit ourselves to take into account only requests that modify the state of the chain (both internal and external), and hence induce transactions, even through \emph{view functions}, namely functions that do not modify the state of the chain may be represented as well. 

%As described above, Ethereum transactions  
The transactions (represented by instances of the class \textit{EthereumTransaction}) induced by interaction requests submitted to  smart contracts are related with instances of the class \textit{EthereumContractInteraction} (subclass of the class \textit{SmartContractInteraction}) by means of the object-property \textsf{describes}. Instances of \textit{EthereumContractInteraction} introduce plan descriptions as in the \ONT{} fashion by means of the object-property \textsf{describes}. Plan descriptions are ways of characterizing requests of actions and  the related actions performed by agents. The most notable subclass of \textit{EthereumContractInteraction} is the class \textit{EtherExchangeSmartContractInteraction}, representing the transferring of Ether cryptocurrency from a wallet to another. The class \textit{EtherExchangeSmartContractInteraction} is also a subclass of the class \textit{CryptocurrencyExchangeBlockchainSmartContractInteraction}, which in turn is a subclass of the class \textit{BlockchainSmartContractInteraction}. 
\end{sloppypar}

An example of representing Ethereum transactions in \ONT{} is illustrated in Appendix A.

\section{The ERC721 protocol in \ONT{}} \label{sec:erc}

\begin{sloppypar}
In this section we show how \ONT{} represents the main ERC721 standard functions for managing non-fungible tokens on the Ethereum blockchain. For space limitations,
we illustrate how the ERC721 standard for minting non-fungible tokens is modelled in \ONT{}, whereas the ERC721 functions for transferring tokens, burning tokens, granting ownership of single tokens, of all the tokens stored in the wallet, and to verify the ownership of tokens, as represented in \ONT{}, are described in Appendices B, C, D, E, and F, respectively. Finally, ERC721 tokens are described in Appendix G. 

As seen in the previous section, the smart contracts defined for the ERC721 token management are introduced by means of instances of the class \textit{EthereumERC721SmartContractAgent} (subclass of the class \textit{Agent}). In   \ONT{}, agents having  similar behaviors inherit the representation of their behavior from a common template providing general descriptions that may be customized by single agents. For this purpose, \ONT{} provides a template for the ERC721 standard  introduced by the individual \textit{ethereum\_ERC721\_smart\_contract\_behavior\_template}, which describes the behaviors of agents minting, burning, and transferring Ethereum NFTs according to the guidelines of the standard. Other functions admitted by the standard ERC721, namely the \emph{approve} function (delegating wallets for managing single tokens), the \emph{setApprovalForAll} function (delegating wallets for managing all the tokens owned), and \emph{ownerOf} (for retrieving the owner of a given token), are also represented in \ONT{}. 
\end{sloppypar}

\begin{figure}[H]
        \centering
        \includegraphics[scale=0.65]{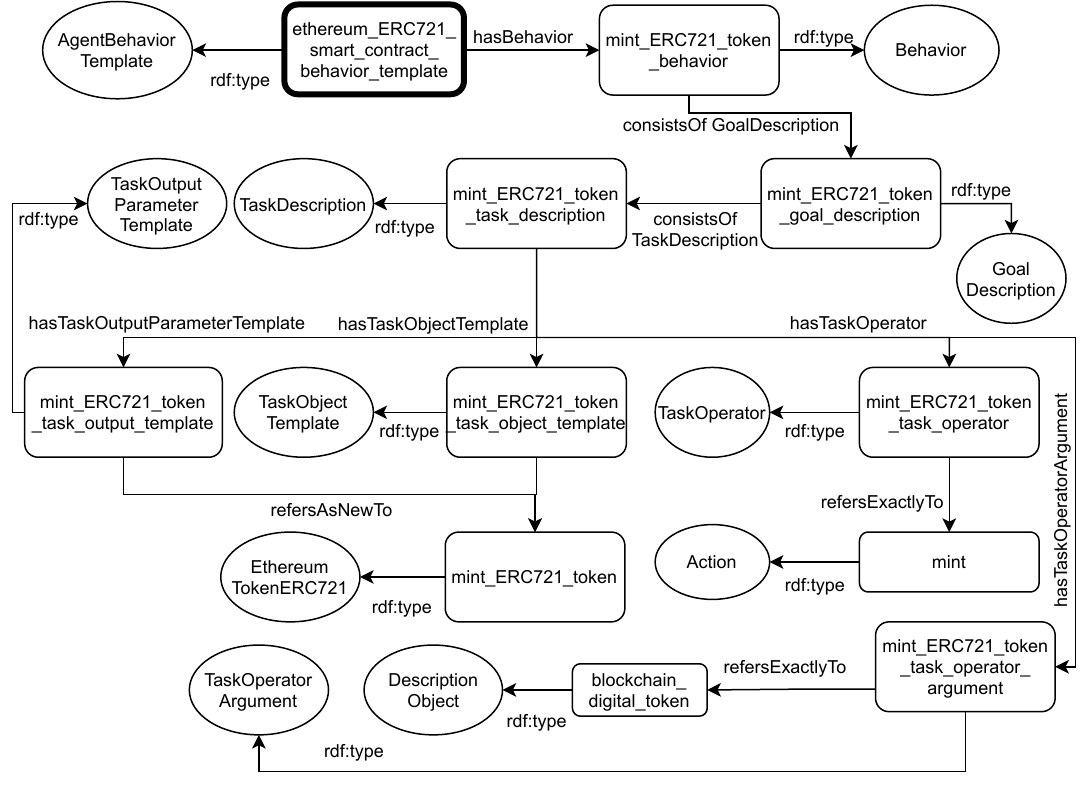}
        \caption{\ONT{} behavior template for the ERC721 minting function}
        \label{fig:mintERC721template}
\end{figure}

Fig.~\ref{fig:mintERC721template} illustrates the behavior template provided by \ONT{} describing the ERC721 function for generating new tokens. The behavior template consists of a single goal, which in turn consists of a single task. Tasks comprise four elements. The first element is the task operator, providing the description of the action to be performed, namely minting. The latter is introduced by means of the individual \textit{mint} (instance of the class \textit{Action}), through the the object-property \textsf{refersExactlyTo}. We recall that in \ONT{} the object-properties \textsf{refersExacltyTo} and \textsf{refersAsNewTo} are introduced to describe the way how constituting elements of agent behaviors must be matched when a  verification of compatible behaviors occurs. The object-property \textsf{refersExactlyTo} introduces well-known entities whose IRIs must correspond to the IRIs of the matched entities or for which the OWL object-property \textsf{sameAs} has been expressed. On the contrary, the object-property \textsf{refersAsNewTo} introduces entities (instances of the class \textit{ReferenceTemplate}) that are used as general descriptions encapsulating the features that the matching entities must satisfy.

The second element of the ERC721 token minting task is the operator argument introducing the individual \textit{blockchain\_digital\_token} by means of the object-property \textsf{refersExactlyTo}. Operators and operator arguments identify unambiguously that the referred operation consists in the generation of (digital) tokens on the blockchain. The third and the fourth elements represent the recipient and the outcome of the operation, respectively. The recipient is introduced by means of a  template of the task object, whereas the outcome is introduced by means of a template of the output parameter of the task. The object template and the output parameter are both connected through the object-property \textsf{refersAsNewTo} to the entity \textit{mint\_ERC721\_token}, which describes the features that the recipient of the action must have, i.e., being an instance of the class \textit{EthereumTokenERC721}.

\section{Conclusions}\label{sec:conclusion}

This paper leveraged the \ONT{} ontology towards the representation of the Ethereum blockchain and the smart contracts deployed on it. Specific focus was on those that comply with the ERC721 standard for NFTs management. \ONT{} is exploited as a means of semantically representing Ethereum  with the aim of probing the blockchain for locating smart contracts and related NFTs by specifying the desired features. 
%In particular, the ontological representation of \ONT{} allows one to find smart contracts and related NFTs, upon the bases of their behavioural descriptions, by invoking purposely crafted SPARQL queries.
%
In particular, the ontological representation of \ONT{} allows one to find smart contracts and related NFTs by inspecting their behavioural descriptions through purposely crafted SPARQL queries.
It was already clear that the \ONT{} approach to semantic representation had the power of generality but our findings demonstrate it at an applied level.

Future work is dense. The very next step is to extend \ONT{} so as to model the standard protocols ERC20 and ERC1155 for fungible and semi-fungible tokens, respectively, and to represent different blockchains such as Stellar Lumens. 
%
%This goal appears to be reachable through the inclusion of endurant and perdurant features.
%
Moreover, we intend to take up the design of a search engine exploiting \ONT{} to find desired smart contracts and tokens using a mechanism of auto-generating parametric \emph{ad-hoc} SPARQL queries that could be borrowed from sibling applications~\cite{woa2019}. 
The present work supports the claim that the potential of the semantic representation of blockchains has much to be unveiled in the near future.

\bibliographystyle{splncs04}

\bibliography{paper}
\clearpage
\section*{Appendix A - Example of Ethereum transaction in \ONT{}}

\begin{figure}[H]
        \centering
        \includegraphics[scale=0.6]{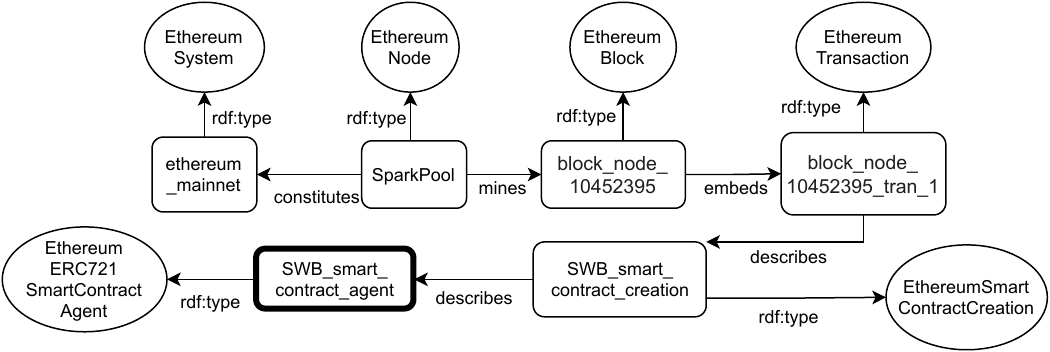}
        \caption{Example of representing an Ethereum transaction in \ONT{}}
        \label{fig:blockOntoexmple}
\end{figure}

\begin{sloppypar}
Fig. \ref{fig:blockOntoexmple} shows an Ethereum transaction storing the smart contract for emitting NFTs  of the Sicilian Wheat Bank (SWB) S.p.A.\footnote{https://www.bancadelgrano.it/en/} The individual \textit{block\_node\_10452395\_tran\_1} represents the Ethereum transaction and is related with the individual \textit{SWB\_SmartContractCreation} describing the smart contract by means of the object-property  \textsf{describes}. The description of the agent identifying the SWB smart contract  is introduced by the individual \textit{SWB\_smart\_contract\_agent} which is linked to \textit{SWB\_smart\_contract\_creation}   by means of  the object-property  \textsf{describes}. Finally, the node (\textit{SparkPool}) that mines the block (\textit{block\_node\_10452395}) including the transaction is also associated to the Ethereum main-net (\textit{ethereum\_mainnet}) by means of the object-property \textsf{constitutes}.
\end{sloppypar}

\clearpage
\section*{Appendix B - ERC721 transferring function in \ONT{}}

\begin{figure}[H]
        \centering
        \includegraphics[scale=0.6]{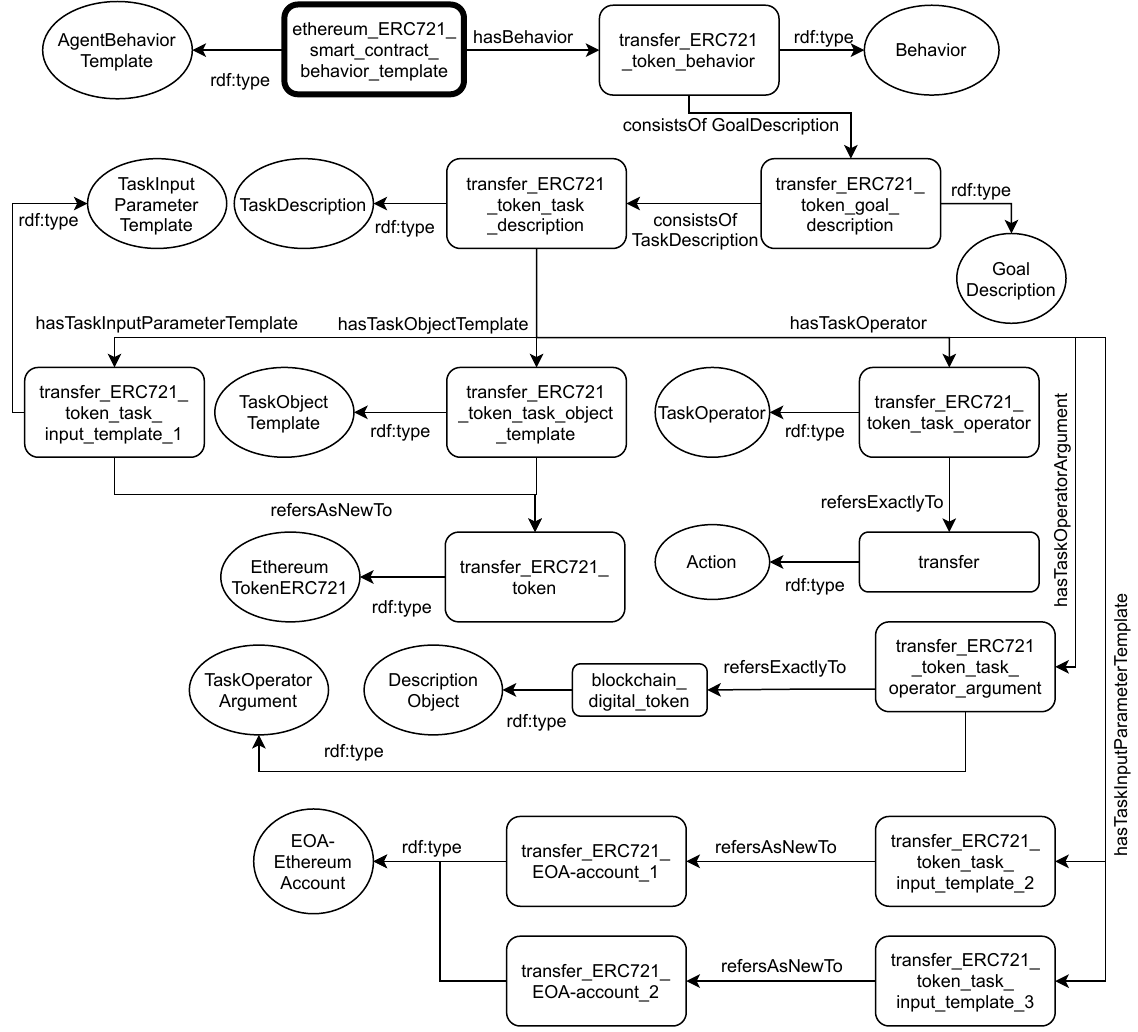}
        \caption{\ONT{} behaviour template for the ERC721 transfer function}
        \label{fig:transferERC721template}
\end{figure}

\begin{figure}[H]
        \centering
        \includegraphics[scale=0.73]{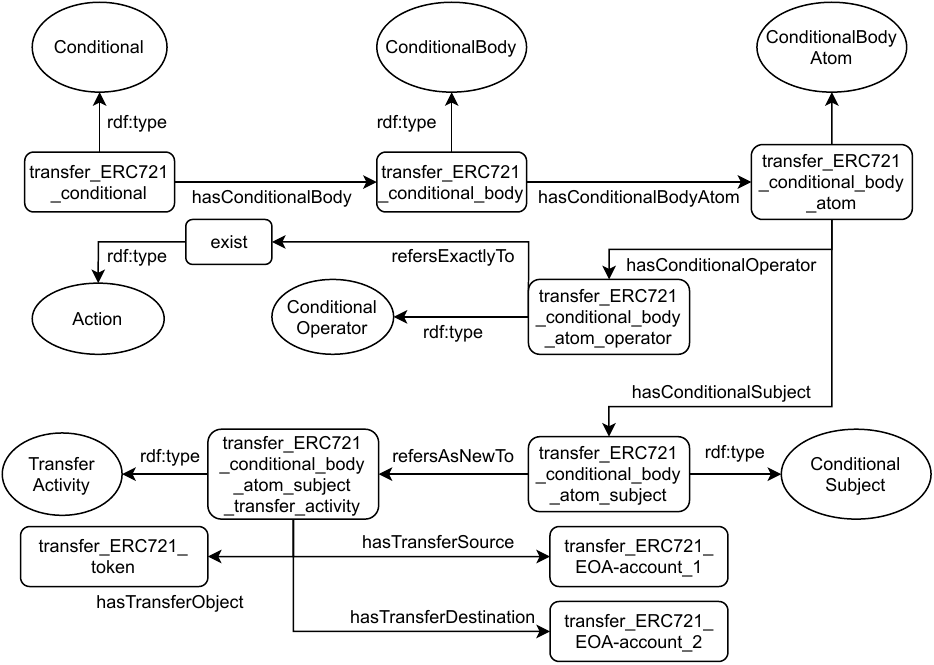}
        \caption{Conditional for the ERC721 transfer function}
        \label{fig:transfercondERC721template}
\end{figure}

\begin{sloppypar}
The behaviour for the ERC721 function for transferring tokens is depicted in Fig. \ref{fig:transferERC721template}. It provides three input parameters, one for the token to be transferred and one for each externally owned account involved in the transferring of the token, namely the source wallet and the destination wallet. The source and destination wallets are introduced by exploiting  the object-property \textsf{refersAsNewTo} by means of two individuals instances of the class \textit{EOA-EthereumAccount}, namely, \textit{transfer-2\_ERC721\_EOA-account} (the source) and \textit{transfer-3\_ERC721\_EOA-account} (the destination). To ensure that the token is transferred from the wallet identified as source to the wallet identified as destination, the conditional illustrated in Fig. \ref{fig:transfercondERC721template} is provided.
\end{sloppypar}

The conditional ensures the existence of a transfer activity for each token to be transferred. The conditional has as conditional object a fresh transfer activity and as operator the individual \textit{exist}. In its turn, the transfer activity indicates as transfer source the wallet used as first parameter in the token transferring, as transfer destination the wallet used as second parameter, and as transferred object the token passed as input parameter.

\clearpage

\section*{Appendix C - ERC721 token burning function  in \ONT{}}

\begin{figure}[H]
        \centering
        \includegraphics[scale=0.65]{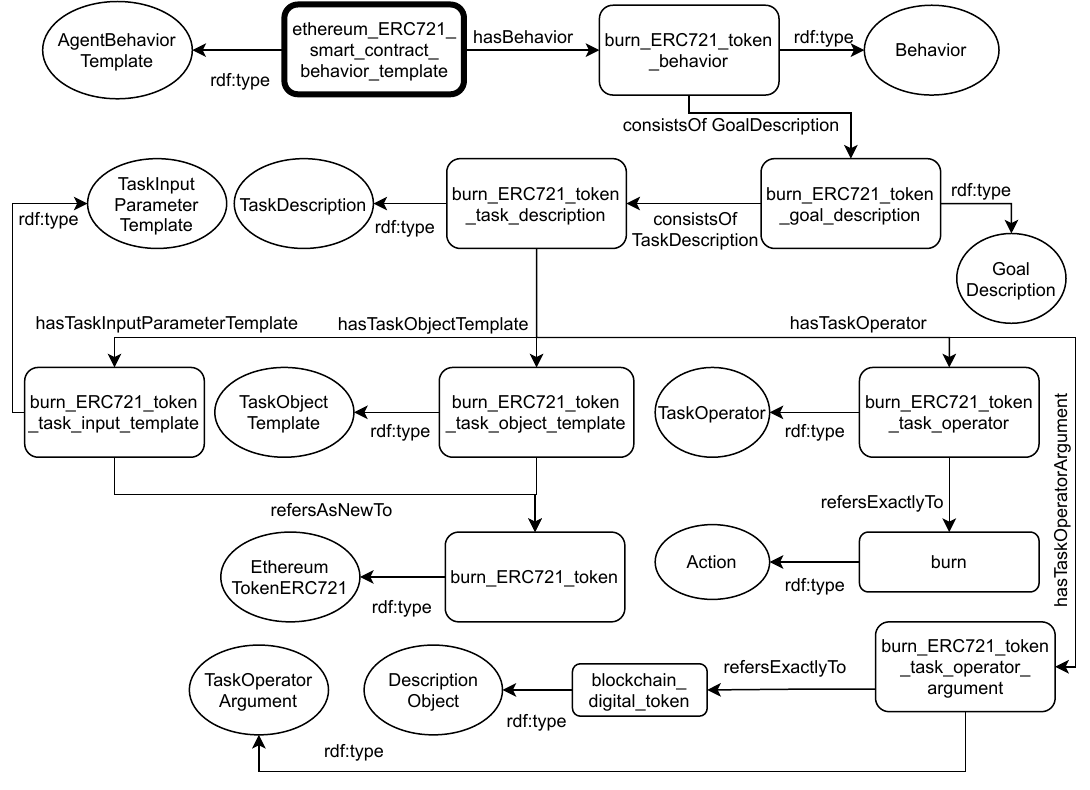}
        \caption{\ONT{} behaviour template for the ERC721 burning function}
        \label{fig:burnERC721template}
\end{figure}

\begin{sloppypar}
The behaviour for the burning function of the ERC721 standard is depicted in Fig.~\ref{fig:burnERC721template}. In this case, the individual \textit{ethereum\_ERC721\_smart\_contract\_behavior\_template} (representing the ERC721 behaviour template) is also connected to the behaviour describing the burning function, whose structure is analogous to the minting function but with a different action and with an input parameter template instead of an output parameter template. The action introduced in the burning function is the individual \textit{burn}, whereas the input parameter template is connected with an individual representing the token to be burnt that is passed as input to the burning function.
\end{sloppypar}

\clearpage

\section*{Appendix D - ERC721 token approve function  in \ONT{}}

The standard ERC721 allows the owner of tokens to delegate wallets to manage tokens on his behalf. Authorizations may be carried out either on a single token or on any token stored in his wallet. The ERC721 function \textsf{approve} and \textsf{setApprovalForAll} are introduced for the former and the latter case, respectively.

\begin{figure}[H]
        \centering
        \includegraphics[scale=0.63]{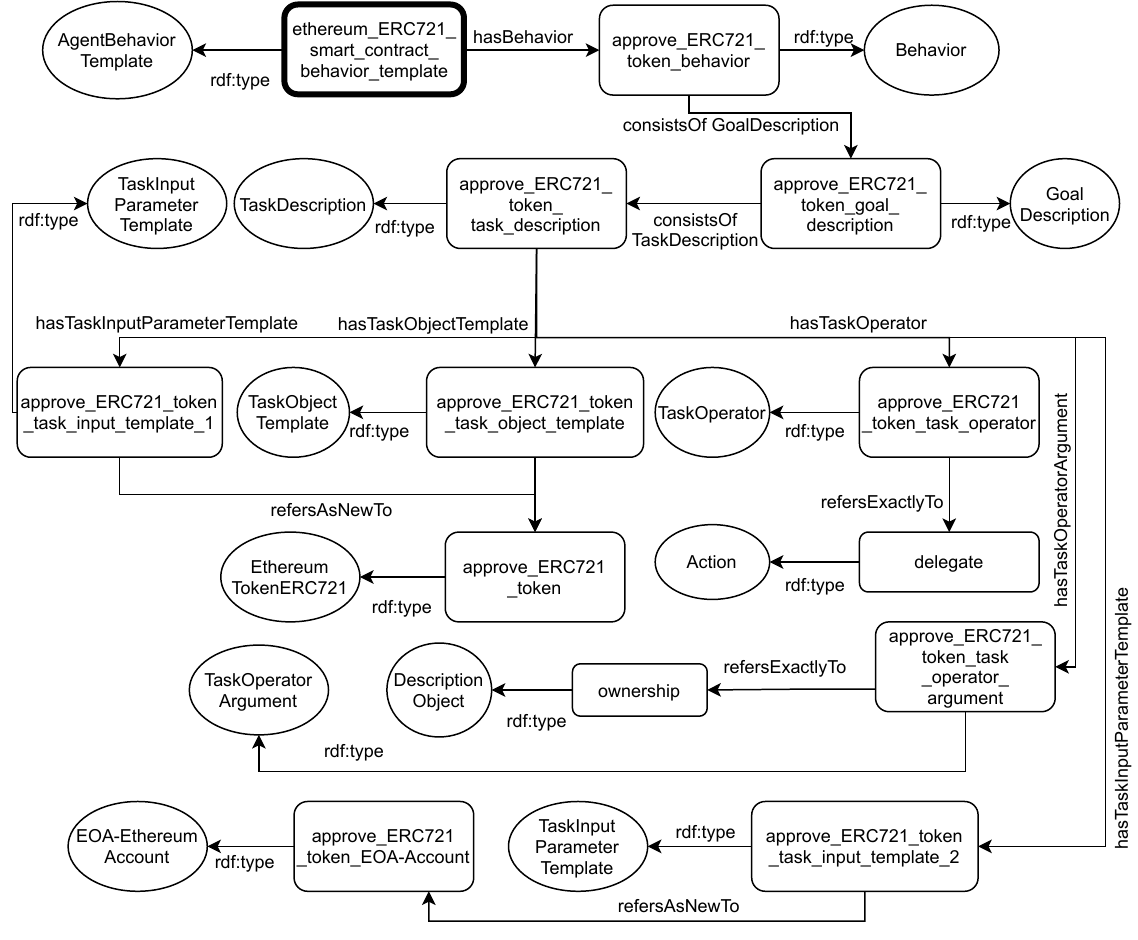}
        \caption{\ONT{} behaviour template for the ERC721 approve function}
        \label{fig:approveERC721template}
\end{figure}

\begin{figure}[H]
        \centering
        \includegraphics[scale=0.65]{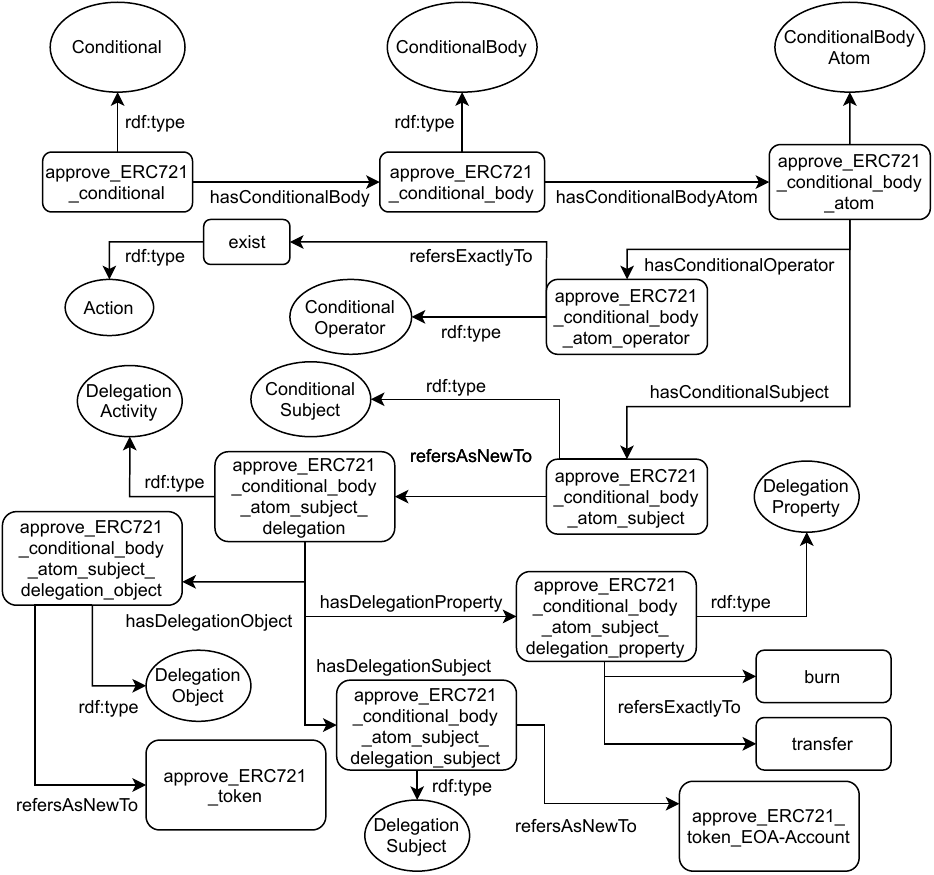}
        \caption{Conditional for the ERC721 approve function}
        \label{fig:approvecondERC721template}
\end{figure}

In case  that an externally owned account is authorized to manage a single token, the behaviour in Fig. \ref{fig:approveERC721template} is adopted. Specifically, the behaviour introduces as input parameter the account to be authorized and the granted token, as operator the instance \textit{delegate}, and as operator argument the instance \textit{ownership}. The conditional in Fig. \ref{fig:approvecondERC721template} ensures that only the operation of burning and transferring may be pursued when the wallet is authorized to operate on behalf of his owner. Such a condition is guaranteed by the existence of a delegation activity (instance of \textit{DelegationActivity}), having as delegation property the instances \textit{burn} and \textit{transfer}. The subject and the object of the delegation activity are the authorized wallet and the granted token, respectively.

%token approve start

%token approve end

\clearpage
\section*{Appendix E - ERC721 setApprovalForAll function in \ONT{}}

\begin{figure}[H]
        \centering
        \includegraphics[scale=0.6]{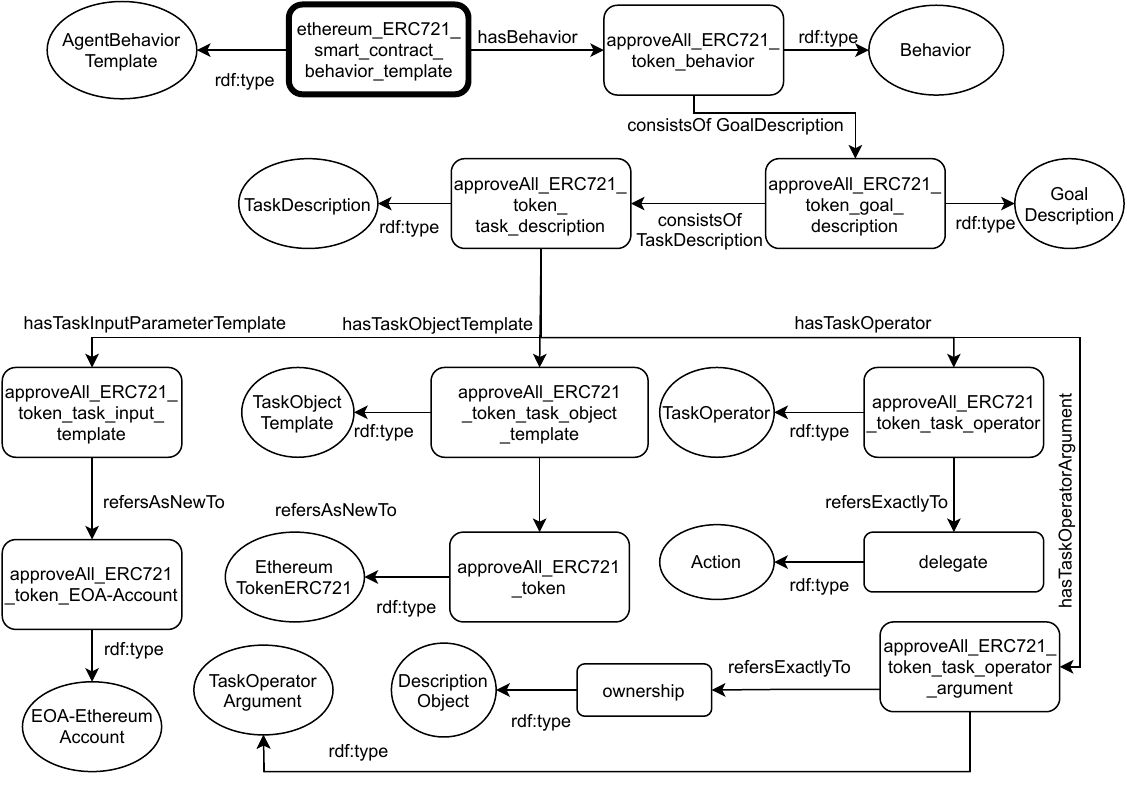}
        \caption{\ONT{} behaviour template for the ERC721 \emph{setApprovalForAll} function}
        \label{fig:approveallERC721template}
\end{figure}

\begin{figure}[H]
        \centering
        \includegraphics[scale=0.65]{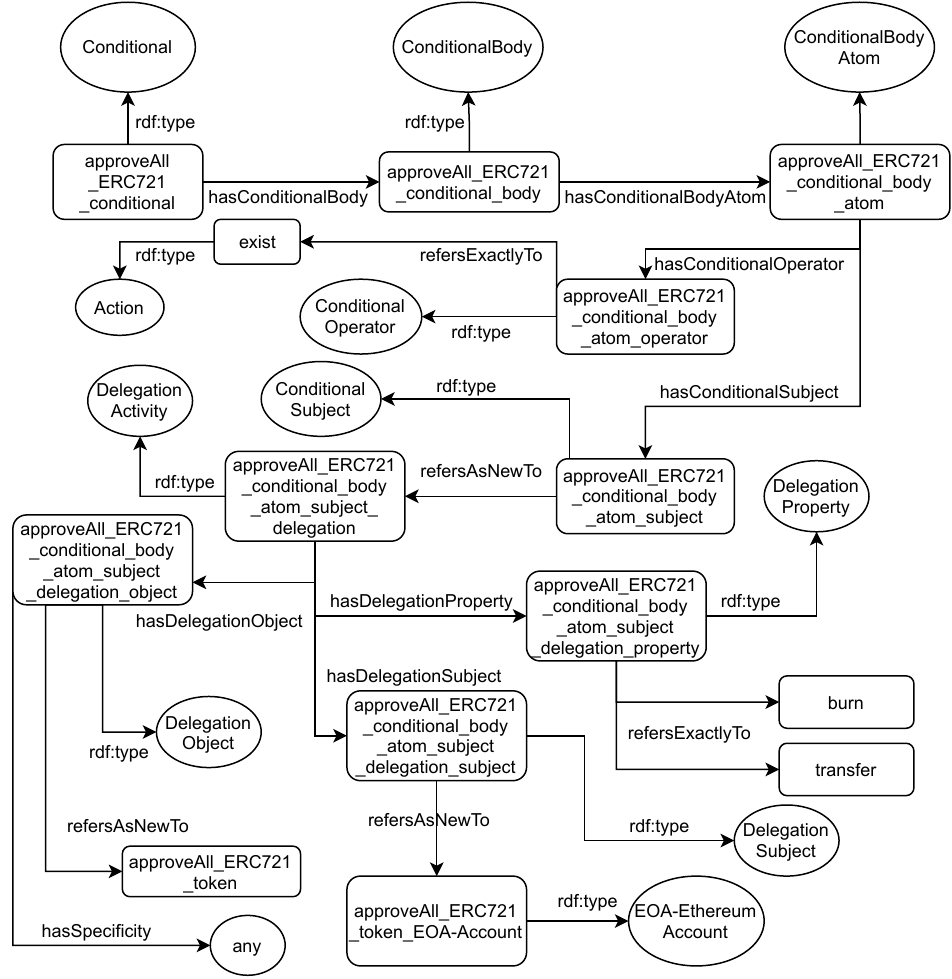}
        \caption{Conditional for the ERC721 approve-all function}
        \label{fig:approveallcondERC721template}
\end{figure}

The Ethereum ERC721 standard allows wallet owners to authorize an account to manage all the tokens owned. The corresponding behaviour is illustrated in Fig.~\ref{fig:approveallERC721template}. The behaviour is close to the behaviour for granting a single token, with the only difference that there is no input parameter concerning the granted token. In such a case, the conditional in Fig. \ref{fig:approveallcondERC721template} ensures that authorization is extended to each token owned. Indeed, the delegation object expresses the object-property \textit{hasSpecificity} with value the individual \textit{any}.

%token approve all start
\clearpage
%token approve all end

\clearpage
\section*{Appendix F - ERC721 token ownership retrieval function  in \ONT{}}

\begin{figure}[H]
        \centering
        \includegraphics[scale=0.62]{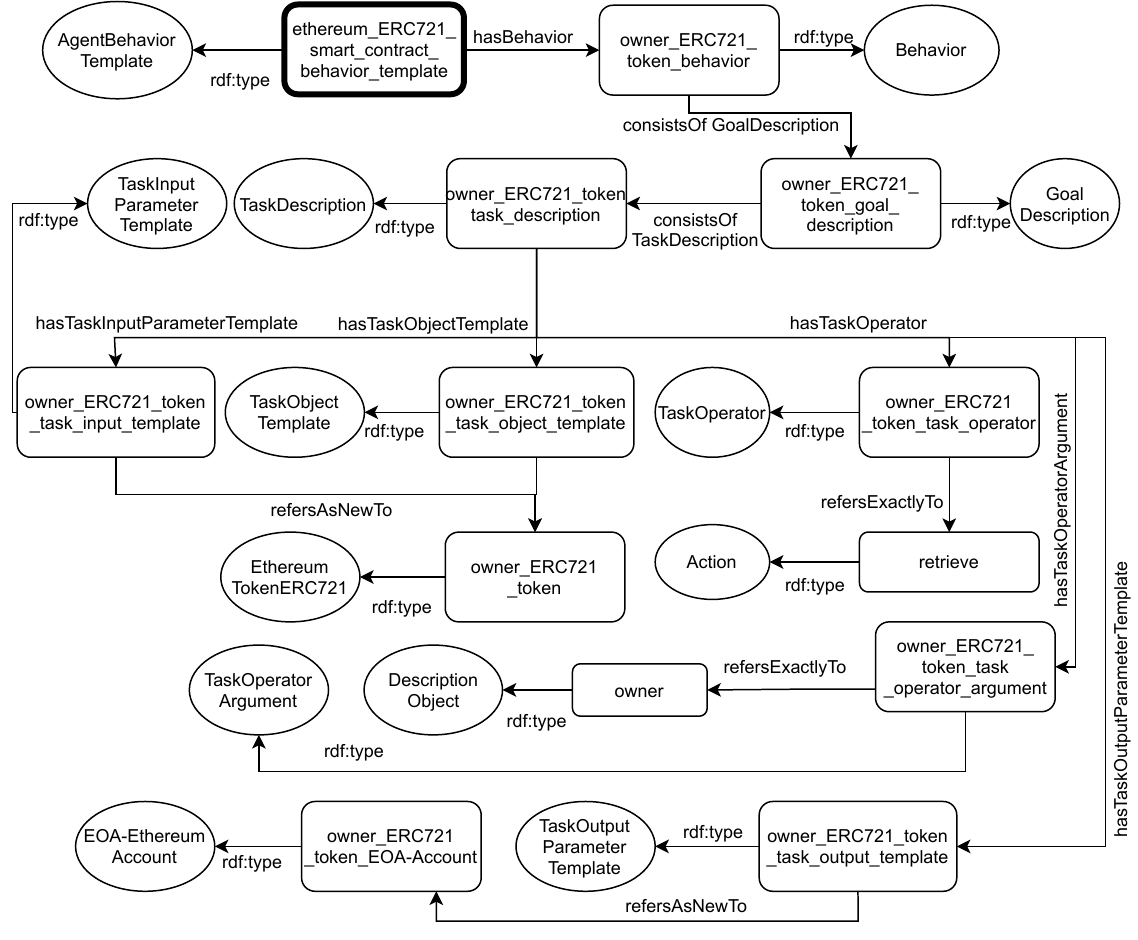}
        \caption{\ONT{} behaviour template for the ERC721 function for retrieving the token's owner}
        \label{fig:ownerofERC721template}
\end{figure}

 The Ethereum ERC721 standard allows one to retrieve the wallet owner of a token. The corresponding behaviour is illustrated in Fig. \ref{fig:ownerofERC721template}. 
 
\begin{figure}[H]
        \centering
        \includegraphics[scale=0.8]{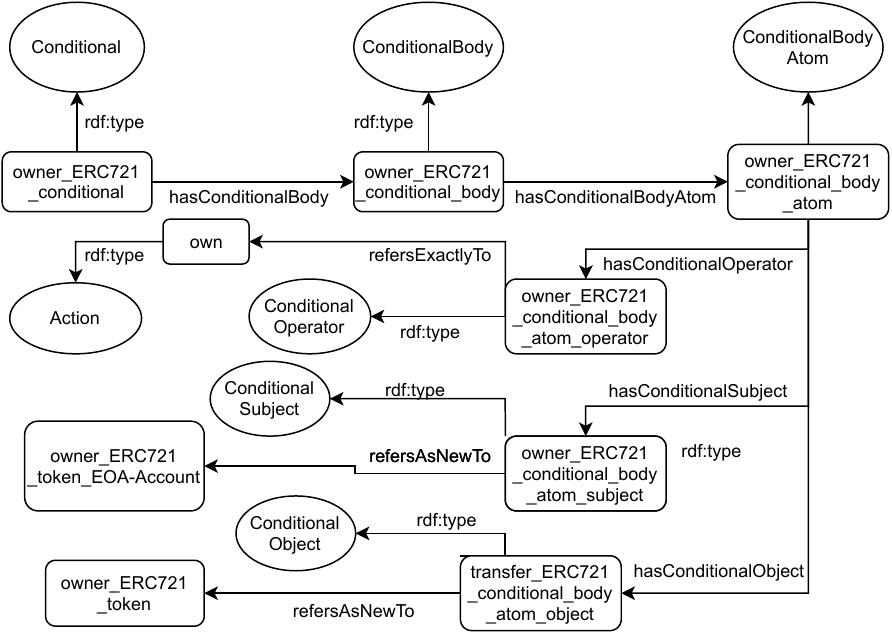}
        \caption{\ONT{} conditional for the ERC721 function for retrieving the token's owner}
        \label{fig:ownerofcondERC721template}
\end{figure}

The  behaviour is related with a task description providing  as recipient and as input parameter the selected token, whose owner has to be retrieved. The conditional in  Fig.~\ref{fig:ownerofcondERC721template} ensures that the wallet retrieved is the effective owner of the token considered.
 
%token owner start
\clearpage
%token owner end

\clearpage
\section*{Appendix G - ERC721 token representation in \ONT{}}

\begin{figure}[H]
        \centering
        \includegraphics[scale=0.8]{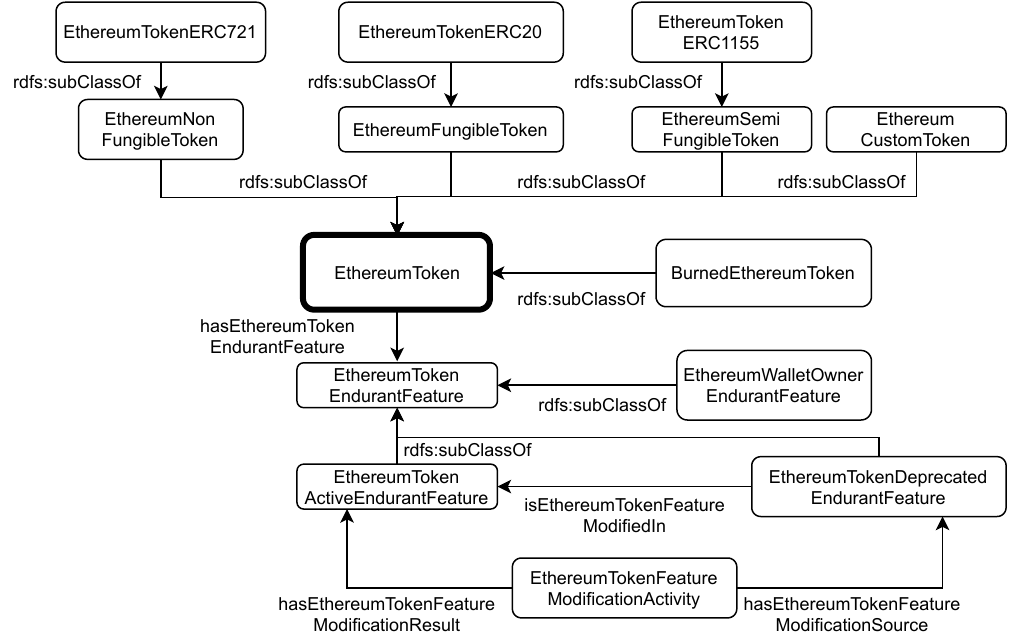}
        \caption{Ethereum token representation in \ONT{}}
        \label{fig:token}
\end{figure}

Tokens are depicted in Fig. \ref{fig:token}. There are four main types of token: 

\begin{itemize}
    \item non-fungible tokens, represented by instances of the class \textit{EthereumSemiFungibleToken}, the latter containing the class \textit{EthereumTokenERC721} that represents non-fungible tokens compliant with the ERC721 standard protocol;
    \item fungible tokens, represented by instances of the class \textit{EthereumFungibleToken}, the latter containing the class \textit{EthereumTokenERC20} that represents fungible tokens compliant with the ERC20 standard protocol;
    \item semi-fungible tokens, represented by instance of the class \textit{EthereumSemiFungibleToken}, the latter containing the class \textit{EthereumTokenERC1155} that represents semi-fungible tokens compliant with the ERC1155 standard protocol;
    \item custom user-defined tokens not compliant with the ERC standard protocols, represented by instances of the class \textit{EthereumCustomToken}.
\end{itemize} 

The four mentioned classes are defined as subclass of the class \textit{EthereumToken}. Additionally, tokens that have been definitively destroyed are also instances of the class \textit{BurnedEthereumToken}.

Tokens carry two types of features~\cite{gangemi2002dolce}, a) perdurant features such as the token ID, that never change and are embedded with the entity representing the token and b) endurant features, that change during the life-span of the token and are associated with an instance of the class \textit{EthereumTokenEndurantFeatures} (subclass of \textit{EndurantFeature}), by means of the object-property \textit{hasEthereumTokenEndurantFeature}. The most notable subclass of \textit{EndurantFeature} is the class \textit{EthereumWalletOwnerEndurantFeature}, which describes the wallet of the  token's owner (by means of the data-properties \textit{isInTheWalletOf}). When the endurant features of a token are modified by the smart contract managing it, they became deprecated and replaced by a new set of features by means of a modification activity. Those new features are introduced by means of a fresh instance of the class \textit{EndurantFeature}. Modification of tokens is allowed only if it involves endurant features and hence perdurant features cannot change. Endurant features may be replaced with other endurant features by introducing an instance of the class \textit{EthereumTokenFeatureModificationActivity} which is connected with:

\begin{itemize}
    \item the changed endurant feature, which is also instance of  the class \textit{DeprecatedEthereumTokenEndurantFeature}, by means of the object-property \textit{hasEthereumTokenFeatureModificationSource};
    \item the new endurant feature, by means of the object-property \textit{hasEthereumTokenFeatureModificationResult}.
\end{itemize}

Moreover, the modified endurant feature is connected with the endurant feature that replaces it by means of the object-property \textit{isEthereumTokenFeature
ModifiedIn}, whereas the token embedding the features is connected with the new endurant feature by means of the object-property \textit{hasEthereumTokenEndurantFeature} as described above.
\end{document}